# Extremely Large Anisotropy of Effective Gilbert Damping in Half-Metallic $CrO_2$

Liangliang Guo[1], Ranran Cai[1], Zhenhua Zhang[2], Wenyu Xing[1], Weiliang Qiao[1], Rui Xiong[3], Zhihong Lu[2], Xincheng Xie[1,4,5], and Wei Han[1]*

[1] International Center for Quantum Materials, School of Physics, Peking University, Beijing 100871, P.R. China

[2] The State Key Laboratory of Refractories and Metallurgy, Wuhan University of Science and Technology, Wuhan 430081, P.R. China

[3] Key Laboratory of Artificial Micro- and Nano-structures of Ministry of Education, School of Physics and Technology, Wuhan University, Wuhan 430072, P.R. China

[4] Interdisciplinary Center for Theoretical Physics and Information Sciences (ICTPIS), Fudan University, Shanghai 200433, P.R. China

[5] Hefei National Laboratory, Hefei 230088, P.R. China

*Correspondence to: weihan@pku.edu.cn

**ABSTRACT**

   Half-metals are a class of quantum materials with 100% spin-polarization at the Fermi level and have attracted a lot of attention for future spintronic device applications. $CrO_2$ is one of the most promising half-metal candidates, for which the electrical and magnetic properties have been intensively studied in the last several decades. Here, we report the observation of a giant anisotropy (~1600%) of effective Gilbert damping in the single crystalline half metallic (100)-$CrO_2$ thin films, which is significantly larger than the values observed on conventional ferromagnetic Fe and CoFe thin films. Furthermore, the effective Gilbert damping exhibits opposite temperature-dependent behaviors below 50 K with magnetic field along [010] direction and near [001] direction. These experimental results suggest the strong spin-orbit coupling anisotropy of the half-metallic $CrO_2$ and might pave the way for future magnonic computing applications.

Anisotropy is one of the most interesting behaviors observed in condensed matter and materials sciences. The electrons, phonons, photons, and spins have been reported to exhibit anisotropic transport properties in single crystalline materials[1-8]. For magnetism, the magnetization anisotropy and anisotropic magnetoresistance[9-11] are extremely important parameters for the magnetic hard disk storage applications[12,13]. Gilbert damping is another key parameter in magnetic storage and computing devices, which quantitatively describes the energy relaxation rate of magnetization dynamics[14,15]. Recent studies have shown the strong anisotropic Gilbert damping in conventional 3d-ferromagnetic metallic films[16,17], with the possible origins from interfacial/bulk spin-orbit couplings compared to the 3d-ferromagnetic metals.

Half-metals are a class of quantum materials with 100% spin polarization[18,19]. With only one spin-polarization at the Fermi level, half-metals are naturally ideal spin sources and detectors. Among all possible half-metal material candidates, $CrO_2$ is the most promising one that has been experimentally confirmed from Andreev reflection measurements[20]. For these reasons, the transport and magnetic properties of $CrO_2$ have been intensively studied[21–26]. In a recent work, an ultralow damping of $CrO_2$ has been reported[27], which is quite intriguing for magnon-based spintronics.

Here, we report ~1600% anisotropy of effective Gilbert damping in a 190 nm single crystalline (100)-$CrO_2$ thin film. Interestingly, the opposite temperature-dependent behaviors of the damping along easy and hard axes of $CrO_2$ have been observed, which further enhance $CrO_2$'s value in temperature mediated applications. Our results suggest the strong anisotropic spin-orbit coupling (SOC) of the half-metallic $CrO_2$ and might be important for theorical investigation of magnetic damping mechanisms in unconventional magnetic materials.

The single crystalline $CrO_2$ films (illustrated in Figure 1(a)) were synthesized on (100)-oriented rutile $TiO_2$ substrates using the chemical vapor deposition technique[27]. Prior to the growth, the $TiO_2$ substrates were ultrasonically cleaned in hydrogen fluoride acid for ~ 2 mins. During the deposition, the $TiO_2$ substrates were kept at 390 °C for the growth of $CrO_2$ films from the source materials of $CrO_3$ powders heated at 260 °C. The crystalline structure properties were characterized via X-ray diffraction measurement (Supplementary Materials I and Figure S1), and the thickness of the 190 nm $CrO_2$ film was determined from XRD reflectometry results on a 27 nm $CrO_2$ film grown under the same condition.

The Ferromagnetic Resonance (FMR) measurements were performed using the vector network technique (PNA-L, Keysight N5234B) under -5 dBm microwave power based on a home-made coplanar wave guide with temperature-variable ability in a Quantum Design Physical Properties Measurement System (PPMS). During the measurement, the single crystalline $CrO_2$ film was attached to the coplanar wave guide with insulating silicon paste. The forward complex transmission coefficients ($S_{21}$), referred to the FMR signal intensity, were recorded as a function of the magnetic field sweeping from ~15000 Oe to 0 Oe under various RF frequencies between 2-40 GHz. Figure 1(b) shows the typical FMR results mapping measured on the 190 nm thick $CrO_2$ thin film.

The electrical conductivity and resistivity of $CrO_2$ thin films were measured via DC methods using van der Pauw geometry in the PPMS. During the measurement, a DC bias voltage 0.5 V was applied using a Keithley K2400, the current and voltage are measured using the same K2400 and a Keithley K2002, respectively. The magnetization hysteresis of $CrO_2$ films were measured in Quantum Design Magnetic Properties Measurement System (MPMS) with magnetic field up to $\pm 2$ T and temperature varies from 300 K to 2 K.

The Gilbert damping characterizes the magnetization dynamics of a ferromagnetic material (illustrated in Figure 1(b) inset), which can be described by Landau-Lifshitz-Gilbert equation[28,29],

$$\frac{d\vec{M}}{dt} = -\gamma \vec{M} \times \vec{H}_{eff} + \frac{\alpha}{M_s} \vec{M} \times \frac{d\vec{M}}{dt} \quad (1)$$

where $\alpha$ is Gilbert damping, $\vec{M}$ is the magnetization vector, $\gamma$ is the gyromagnetic ratio, and $M_s = |\vec{M}|$ is the saturation magnetization.

Figure 1(c) shows the FMR signal measured with the external magnetic field along various magnetic field directions ($\varphi_H$) at $T$ = 300 K and $f$ = 25 GHz. These experimental results could be fitted using the Lorentz equation (solid lines in Figure 1(c))[30]:

$$S_{21} \propto S_0 \frac{(\Delta H)^2}{(\Delta H)^2 + (H - H_R)^2} \quad (2)$$

where $S_0$ is the constant describing the coefficient for the transmitted microwave power, $H$ is the external magnetic field, $H_R$ is the magnetic field under the resonance condition, and $\Delta H$ is the half linewidth (labelled in Figure 1(c)).

Clearly, as the magnetic field direction varies, a large difference of the FMR signal is noticed. Firstly, one obvious feature is that as the magnetic field direction rotates away from the $CrO_2$'s [010] crystalline direction, the resonance magnetic field ($H_R$) decreases from ~ 6700 Oe to ~ 5800 Oe with the magnetic field along $CrO_2$'s [001] direction. This magnetic anisotropy has also been shown in the magnetization hysteresis measurement (Figure 1(c) inset and Figure S2). The anisotropic $H_R$ is summarized in Figure 1(d), which can be described by the following equation[16],

$$\left(\frac{2\pi f}{\gamma}\right)^2 = \left[H_R \cos(\varphi_M - \varphi_H) + 4\pi M_{eff} - U \sin^2(\varphi_M) + \frac{B}{4}(3 - \cos(4\varphi_M))\right]$$
$$\times [H_R \cos(\varphi_M - \varphi_H) - U \cos(2\varphi_M) - B \cos(4\varphi_M)] \qquad (3)$$

where $U$ is the in-plane uniaxial anisotropy, $B$ is the biaxial or cubic magnetic anisotropy, $4\pi M_{eff}$ is the effective magnetization, and $\varphi_M$ illustrates the magnetization angle with respect to the crystal's [010] direction. Since the measurement is performed under large magnetic fields with $f$ = 25 GHz, we can steadily assume $\varphi_M$ is very close to $\varphi_H$ and the best fitting is shown as the black curve in Figure 1(d). Secondly and more interestingly, the resonance dip becomes sharper and the linewidth dramatically decreases as the magnetic field change from $CrO_2$'s [010] to the [001] crystalline direction. The anisotropic $\Delta H$ is summarized in Figure 1(e), where a ~5 times difference of $\Delta H$ is observed.

To quantitatively obtain the effective Gilbert damping, the FMR signal is measured at various frequencies, and frequency-dependent half linewidth is analyzed. As shown in Figure 2(a), the extracted $\Delta H$ vs. $f$ is plotted for the results measured with magnetic field direction along CrO$_2$'s [010] direction and $\varphi_H$ = 60º. The effective Gilbert damping can be obtained from the linearly fitted curves (solid lines in Figure 2(a)) of the $\Delta H$ vs. $f$ in the high-frequency linear region using the following equation,

$$\Delta H = \left(\frac{2\pi}{\gamma}\right)\alpha f + \Delta H_0 \qquad (4)$$

where $\Delta H_0$ is related to the inhomogeneous properties of the CrO$_2$ films. The effective Gilbert damping with magnetic field direction along CrO$_2$'s [010] and $\varphi_H$ = 60º is calculated to be 0.0088 ± 0.0004 and 0.0023 ± 0.0002 respectively. Another way to obtain the damping is called Low-field losses fitting [31,32], from which similar damping values are obtained (Supplementary Materials II and Figure S3). When the external magnetic field is applied along CrO$_2$'s [001] direction, $\Delta H$ keeps decreasing as $f$ increases up to 40 GHz (Figure 2(b)), which makes it impossible to obtain reliable effective Gilbert damping using eq. (4) or using the Low-field losses fitting (Figure S4). Thus, we only discuss the obtained effective Gilbert damping values up to $\varphi_H$ = ±60º, which are reliable using both linear fitting and Low-field losses fitting.

The obtained effective Gilbert damping is summarized in Figure 2(c) as the magnetic field angle varies from $\varphi_H$ = -60º to $\varphi_H$ = 60º. When the magnetic field is along the CrO$_2$'s [010] direction ($\varphi_H$ = 0º), the largest damping is observed 0.0088 ± 0.0004. And when the magnetic field angle rotates away from the CrO$_2$'s [010] direction, the effective Gilbert damping decreases quickly, and the lowest damping is observed to be 0.0023 ± 0.0002 and 0.0022 ± 0.0002 for $\varphi_H$ = 60º and -60º, respectively. Although the damping values with $H$ along CrO$_2$'s [001] direction cannot be determined, we strongly believe those will be much smaller given the fact that the half linewidths measured with $H$ along CrO$_2$'s [001] direction are already smaller than the values of $\varphi_H = 60°$ (Figure 1(e)). The difference is quite large for ~ 4 times for effective Gilbert damping between $\varphi_H = 0°$ and $\varphi_H = 60°$, which is considerably larger compared to ultrathin Fe films on GaAs substrates[16]. If we consider $\varphi_H = 90°$, the anisotropy would be much larger. For ultrathin Fe films on GaAs substrates, the interface symmetry breaking generates Rashba SOC, which is the explanation for the emergence of anisotropy of effective Gilbert damping[16]. However, for the single crystalline $CrO_2$ in this report, the $CrO_2$ film is ~ 190 nm, which is extremely thick which means the role of Rashba SOC at the interface between $CrO_2$ thin film and $TiO_2$ substrate would be negligible. Therefore, we believe the anisotropic effective Gilbert damping in $CrO_2$ is the bulk property rather than substrate/interface effect in previous studies of Fe/GaAs[16]. Since the intrinsic origin of the Gilbert damping is believed to arise from spin-orbit coupling[14,33–36], one possible reason of the anisotropic damping is the effective SOC strength depends on the spin orientations in $CrO_2$. For the spins along [001] direction, the effective SOC is small, which results quite small damping constant for such large SOC materials. However, for the [010] direction, the effective SOC is quite strong, leading to quite large damping. Nevertheless, further theoretical calculations are needed to confirm.

To further investigate the large anisotropy, we measure temperature evolution of effective Gilbert damping in $CrO_2$ films. Figure 3(a) shows $\Delta H$ as a function of $f$ with the external magnetic field along the $CrO_2$ crystal's [010] direction at $T = 200, 100, 50$, and 25 K, respectively. Clearly, as the temperature decreases, the slope of $\Delta H$ vs. $f$ increases. Especially for $T = 25$ K, the slope is significantly large compared to the other temperature results. When $T$ is lower than 25 K, the FMR signal is very broad, and the FMR intensity is significantly small that prohibits the accurate determination of the effective Gilbert damping values. Nevertheless, a much broader signal usually indicates a higher effective Gilbert damping. To be noted, the very good linear relation between half linewidth and frequency (up to 40 GHz) can exclude the contribution of two-magnon scattering. The orientation and temperature dependent damping behavior can exclude the contributions of the inhomogeneous and mosaicity broadening terms (See Supplementary Materials IV). Figure 3(b) shows $\Delta H$ as a function of $f$ with the external magnetic field along the $\varphi_H = 60°$ at $T = 200, 100, 50$, and $2$ K, respectively. The temperature-dependent effective Gilbert damping results are summarized in Figure 3(c). Clearly, a larger difference between the effective Gilbert damping in various magnetic field direction is observed. Figure 3(d) compares the anisotropy of effective Gilbert damping in the 190 nm $CrO_2$ film as a function of $\varphi_H$. As the temperature decreases from 300 K to 25 K, the anisotropy strongly increases, and reaches to ~1600% at $T = 25$ K. Although the actual value of effective Gilbert damping for $H$ along [010] direction is not obtained at the temperature below 25 K, an enhanced damping is expected since the FMR signal is much broader for $T < 25$ K. Hence, we strongly believe that the anisotropy will be even larger when the temperature further decreases below $T = 25$ K, which might be resolved in future studies via more sensitive probes.

Let us discuss the physical mechanisms of the observed large anisotropic effective Gilbert damping in the $CrO_2$ thin films. Previous reports show that effective Gilbert damping could exhibit resistivity or conductivity-like feature[37–41]. Next, we investigate this relationship through straightforwardly comparing the temperature-dependent effective Gilbert damping and electrical transport resistivity/conductivity (see Methods). Figures 4(a) and 4(b) contrast the temperature-dependent effective Gilbert damping of $CrO_2$ film with magnetic field along two different directions $\varphi_H = 0°$ and $\varphi_H = 60°$. The insets are the conductivity and resistivity. Clearly, the effective Gilbert damping along [010] direction and $\varphi_H = 60°$ exhibit opposite features while temperature decreases from $T = 50$ K to $T = 2$ K. Such opposite trends of temperature-dependent effective Gilbert damping have never been reported based on our best understanding, for which the full understanding will need future theoretical investigations. Nevertheless, this observation further suggests our speculation that a large anisotropy of SOC strength exists in the single crystalline $CrO_2$. The giant anisotropic effective Gilbert damping has also been reproduced on additional samples, with one of them showing in the Supplementary Materials III and Figure S5.

In summary, a giant anisotropic effective Gilbert damping of ~1600% has been observed in the single crystalline half metallic (100)-$CrO_2$ thin film, which is much larger than those of the conventional ferromagnetic Fe and CoFe thin films in previous reports[16,17]. These large anisotropic properties of effective Gilbert damping could indicate a strong anisotropy of spin-orbit coupling strength of $CrO_2$. These results provide new insights to the damping mechanism in half-metallic $CrO_2$, which might be important for developing future spintronic and magnonic devices using half metals.

**Supporting Information**

XRD, transport and magnetization characterizations, effective Gilbert damping fitted via low-field losses method, additional sample with similar damping behaviors, exclusion of extrinsic contributions to the effective Gilbert damping.

**Author Contributions**

W.H. conceived and supervised the study. L.G. and R.C. performed the magnetization dynamics and the electric conductivity measurement of the $CrO_2$. Z.Z. and R.X. synthesized the single crystalline $CrO_2$ films. All the authors discussed the results and contributed to the final version of the manuscript. L.G. and W.H. wrote the manuscript with contributions from all authors. All the authors discussed the results.

**Notes**

The authors declare no competing financial interest.

**ACKNOWLEDGMENTS**

We acknowledge the financial support from National Basic Research Programs of China (Nos. 2022YFA1405100 and 2019YFA0308401), National Natural Science Foundation of China (NSFC Grant No. 11974025), and the Key Research Program of the Chinese Academy of Sciences (Grant No. XDB28000000).

**Figure 1**

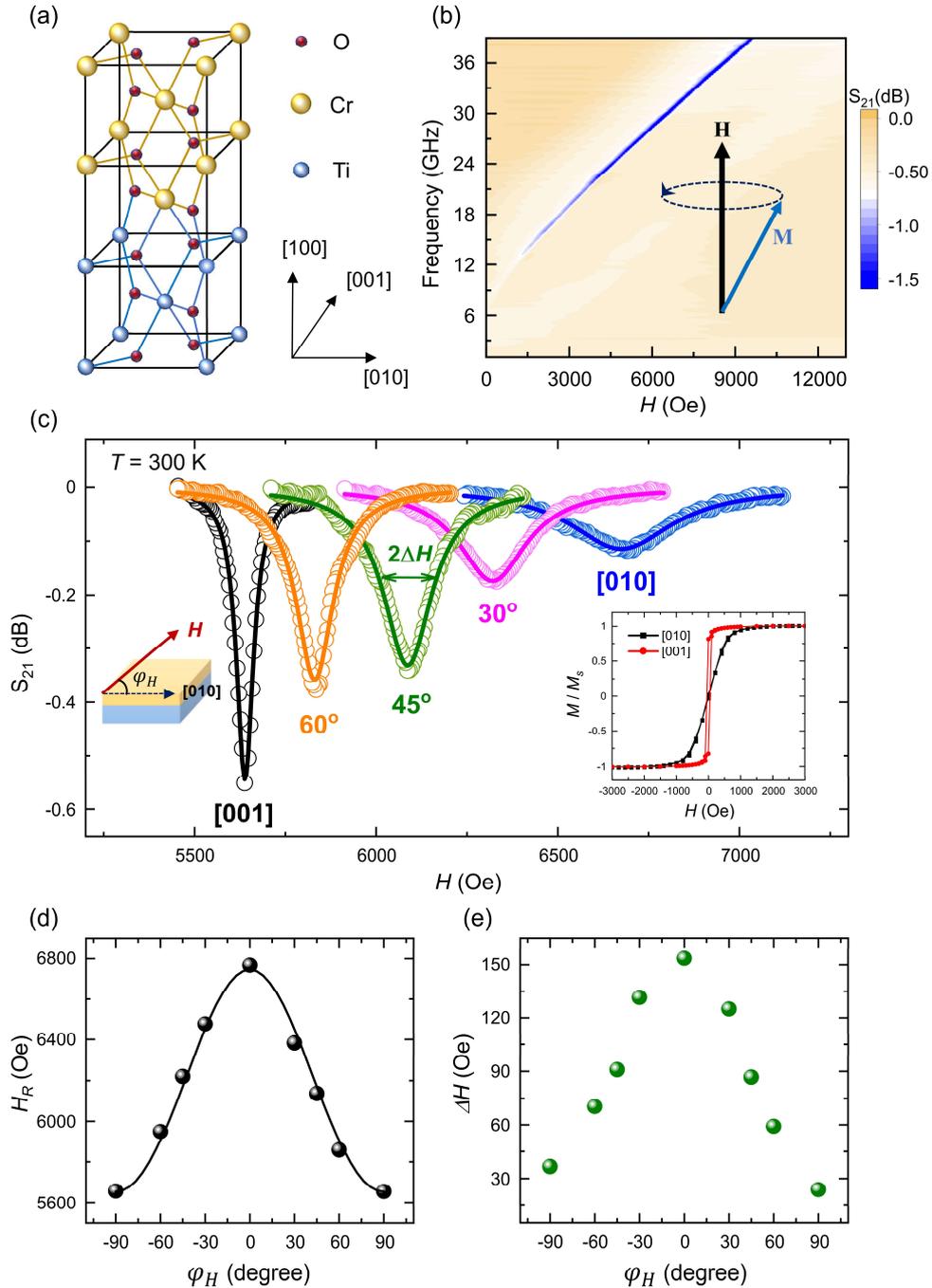

**Figure 1**. Magnetization dynamics of the CrO$_2$ film. (a) Schematic of the crystalline structure of the (100)-oriented CrO$_2$ film grown on rutile TiO$_2$ substrate. (b) Typical FMR absorption map when sweeping frequency at different magnetic field. Inset: a schematic drawing of FMR. (c) The FMR spectra of the $CrO_2$ thin film with the magnetic field along various directions at $T$ = 300 K and $f$ = 25 GHz. A linear background is subtracted for the FMR signal normalization. The solid lines represent the best Lorentz fitting results based on equation (2). Inset: Schematic of the magnetic field angle ($\varphi_H$) during the FMR measurement and the magnetization hysteresis at $T$ = 300 K. (d, e) The resonance magnetic field and half linewidth as a function of $\varphi_H$ measured at $T$ = 300 K and $f$ = 25 GHz.

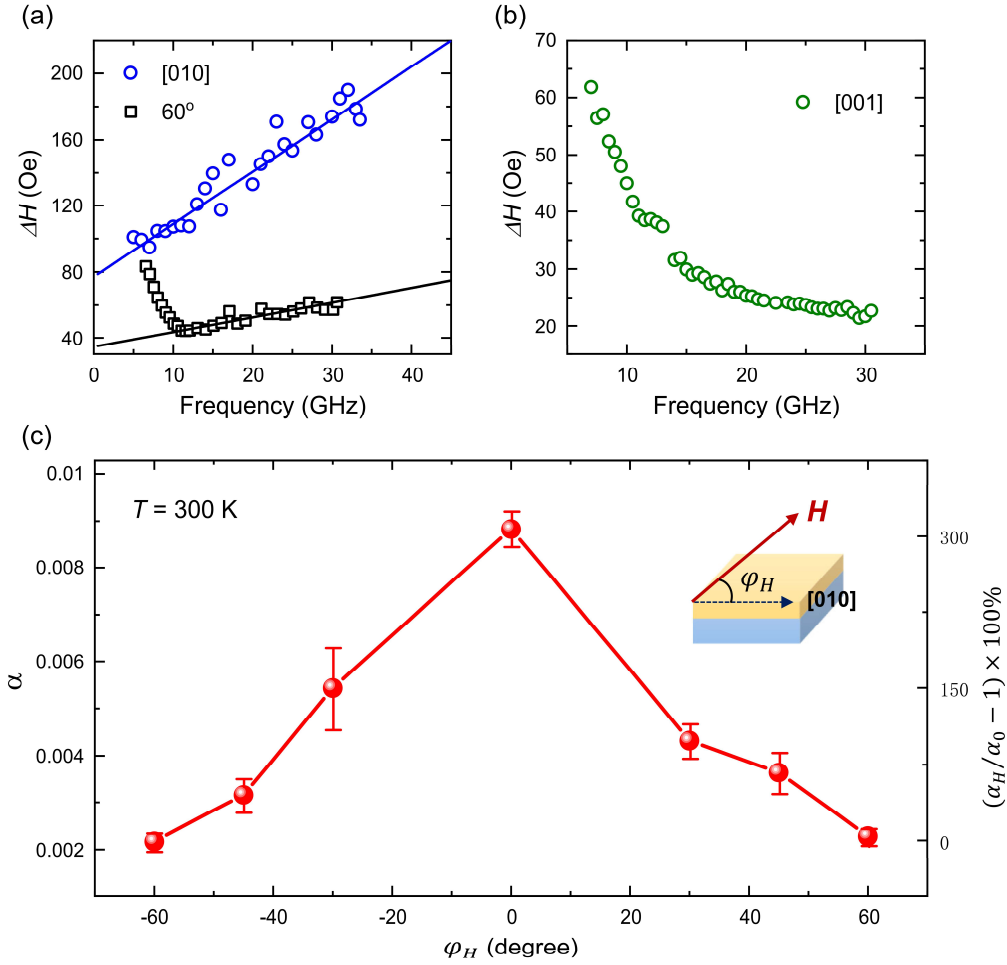

**Figure 2**. Large anisotropy of magnetization dynamics of $CrO_2$ film at $T = 300$ K. (a, b) The resonance half linewidth as a function of RF frequency at $T = 300$ K with the magnetic field along the crystal's [010] direction ($\varphi_H = 0°$) and $\varphi_H = 60°$. The solid lines represent the best fitting curves, from which the effective Gilbert damping values are determined. (c) Large anisotropy of effective Gilbert damping in the $CrO_2$ film as a function of $\varphi_H$ (illustrated in the inset) at $T = 300$ K.

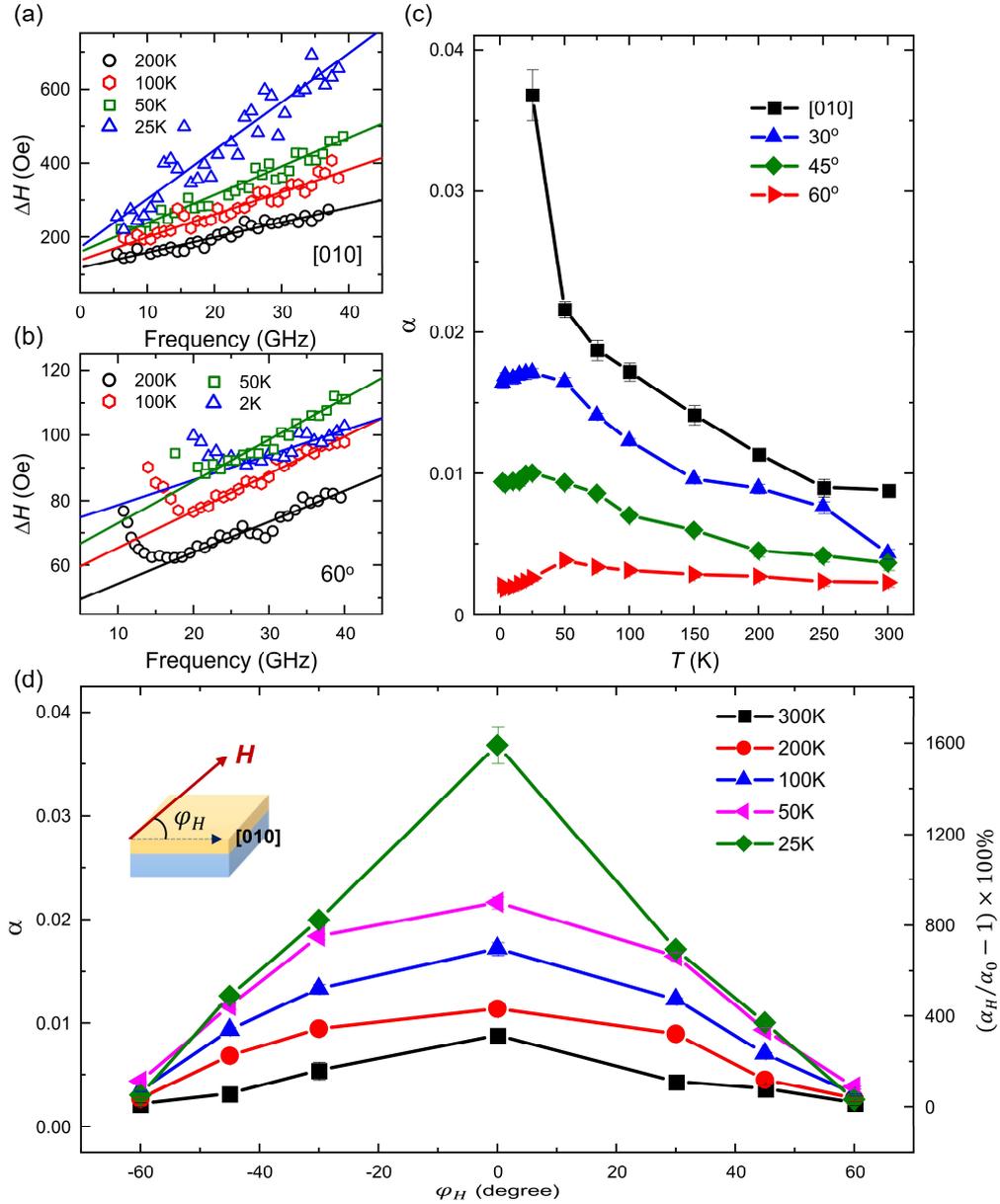

**Figure 3**. Extremely large anisotropy of the effective Gilbert damping of CrO$_2$ at low temperatures. (a, b) The resonance half linewidth as a function of RF frequency with external magnetic field along the CrO$_2$ crystal's [010] direction and $\varphi_H = 60°$ measured at $T$ = 200, 100, 50, and 25 K (2 K), respectively. The solid lines represent the best fitting curves to obtain the effective Gilbert damping. (c) Temperature dependence of the effective Gilbert damping measured

under various $\varphi_H$. (d) Temperature evolution of the anisotropic effective Gilbert damping as a function of $\varphi_H$ (illustrated in the inset).

**Figure 4**

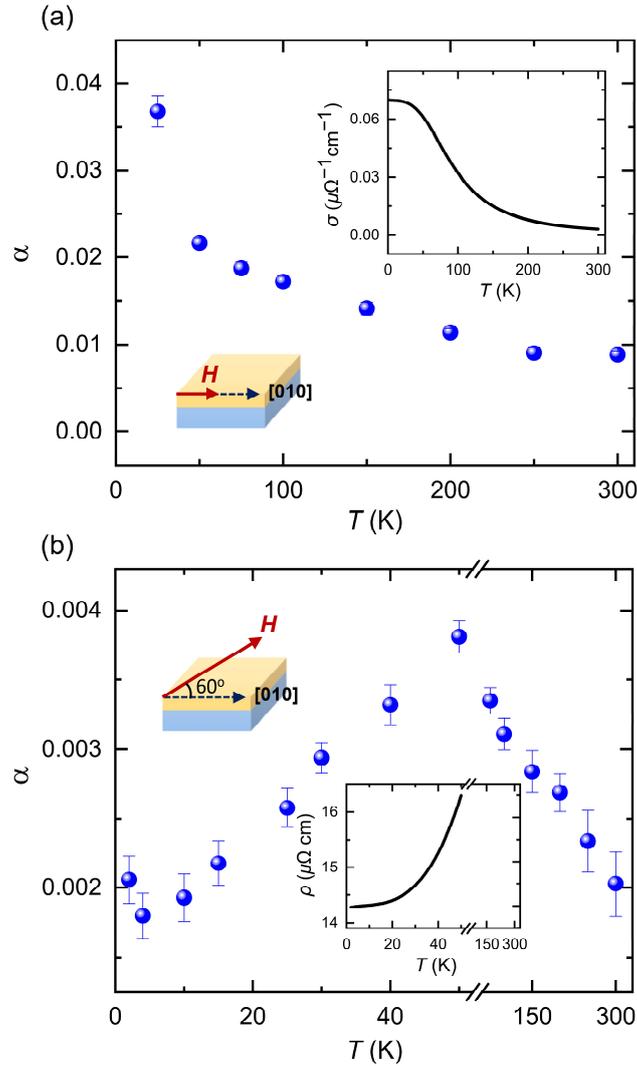

**Figure 4**. Opposite temperature-dependent behaviors of the anisotropic effective Gilbert damping of $CrO_2$. (a, b) Temperature-dependent effective Gilbert damping (blue symbols) when magnetic field is along [010] in (a) and $\varphi_H = 60°$ direction in (b). The insets are the electrical conductivity and resistivity of the $CrO_2$ film.

# Supporting Informatione for:

# Extremely Large Anisotropy of Effective Gilbert Damping in Half-Metallic $CrO_2$

Liangliang Guo[1], Ranran Cai[1], Zhenhua Zhang[2], Wenyu Xing[1], Weiliang Qiao[1], Rui Xiong[3], Zhihong Lu[2], Xincheng Xie[1,4,5], and Wei Han[1]*

[1] International Center for Quantum Materials, School of Physics, Peking University, Beijing 100871, P.R. China

[2] The State Key Laboratory of Refractories and Metallurgy, Wuhan University of Science and Technology, Wuhan 430081, P.R. China

[3] Key Laboratory of Artificial Micro- and Nano-structures of Ministry of Education, School of Physics and Technology, Wuhan University, Wuhan 430072, P.R. China

[4] Interdisciplinary Center for Theoretical Physics and Information Sciences (ICTPIS), Fudan University, Shanghai 200433, P.R. China

[5] Hefei National Laboratory, Hefei 230088, P.R. China

*Correspondence to: weihan@pku.edu.cn

## I: The characterization of the single crystalline (100)-$CrO_2$ film

Figures S1-S2 show the crystalline, electron transport, and magnetic properties of the 190 nm single crystalline (100)-$CrO_2$ film. The XRD results show that only (200)-$CrO_2$ peaks are observed on the rutile (100)-$TiO_2$ substrates (Figure S1(a)), which presents the good single crystalline

properties of the CrO$_2$ film. As temperature decreases, the resistance of the CrO$_2$ film deceases, indicating its good metallic properties (Figure S1(b)).

Figure S2 summarizes the magnetization hysteresis curves with magnetic field along the crystal's [010] and [100] directions at $T$ = 200, 100, 50, and 2 K, respectively. These results together with Figure 1(c) inset ($T$ = 300 K) show that the magnetic anisotropy is along the same direction for all the temperatures from $T$ = 2 to 300 K. This observation is consistent with the magnetic anisotropy results on the single crystalline (100)-CrO$_2$ film in our previous study[1].

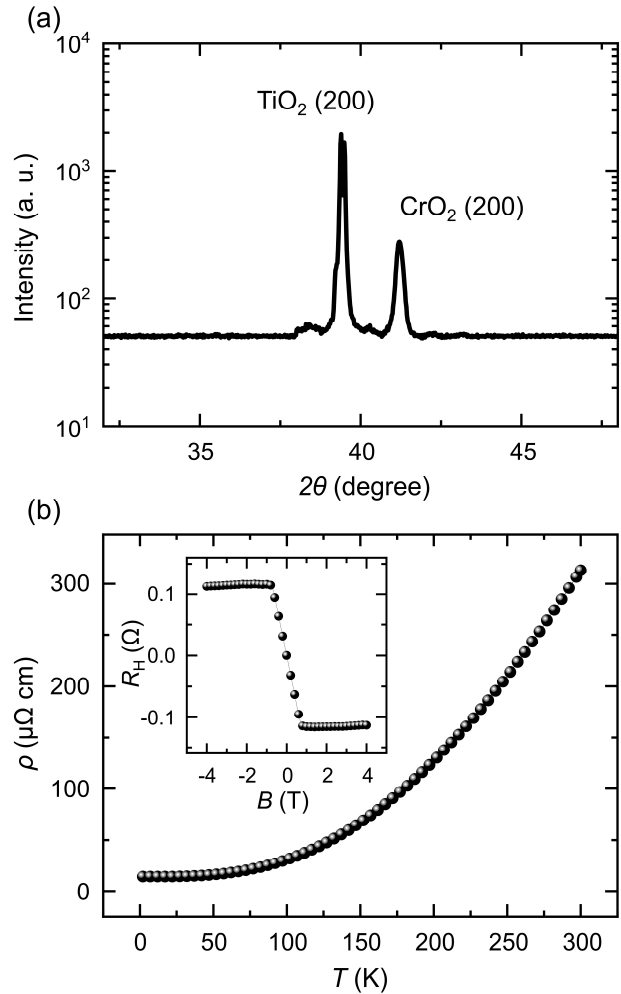

**Figure S1**. Basic characterization of the $CrO_2$ film. (a) XRD results of the 190 nm (100)-oriented $CrO_2$ films grown on rutile $TiO_2$ substrate. (b) The temperature-dependent resistivity of the $CrO_2$ film measured via van der Pauw method. Inset: The anomalous Hall resistance at $T = 300$ K.

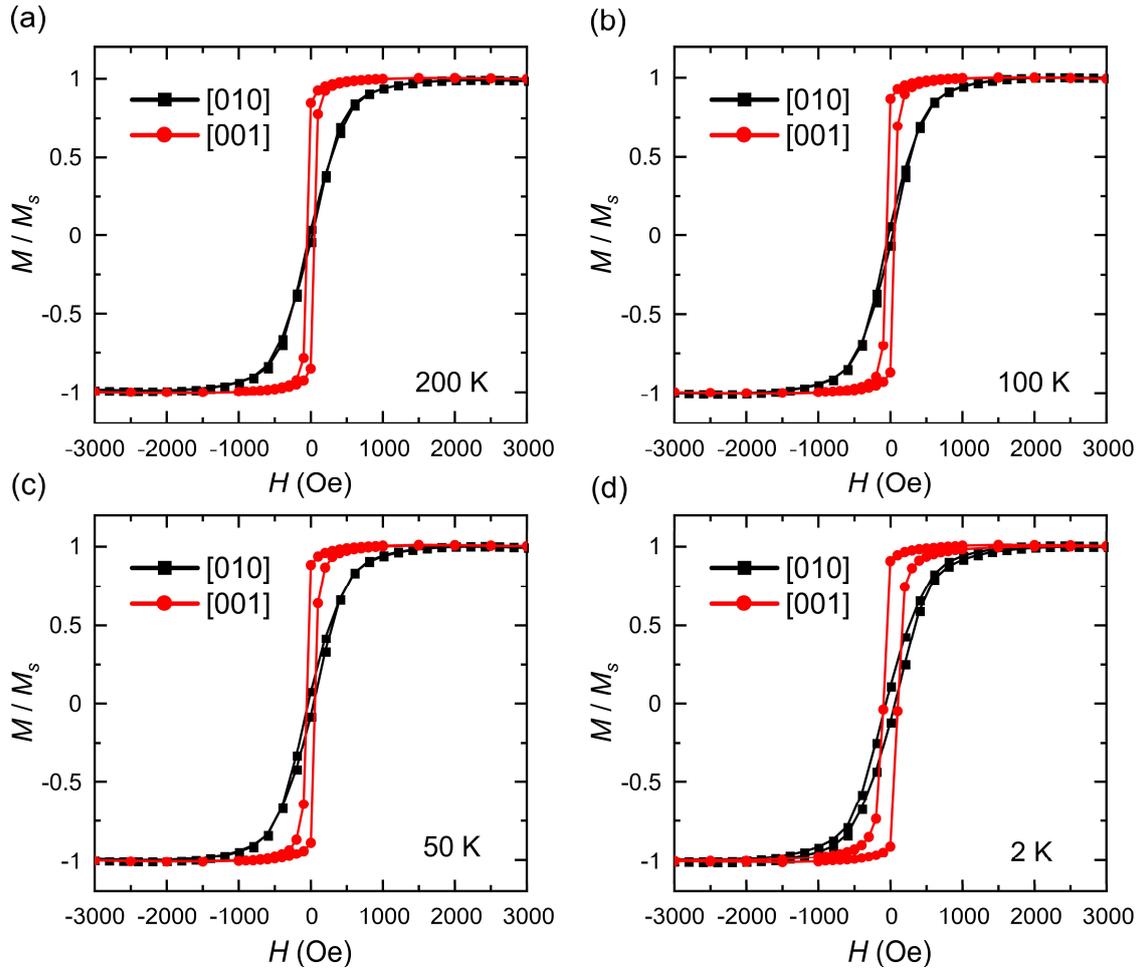

**Figure S2**. Magnetization hysteresis of $CrO_2$ film at low temperatures. (a-d) The magnetization hysteresis loops measured at various temperatures from $T = 200$ to 2 K. The red and black curves represent the measurement with magnetic field along $CrO_2$ crystal's [001] and [010] directions, respectively.

**II: The determination of the effective Gilbert damping via Low-field losses fitting**

For the nonlinear relationship of $\Delta H$ vs. $f$ in the whole frequency range, the Low-field losses fitting has been widely used[2,3], as shown in the equation below:

$$\Delta H = \Delta H_0 + \alpha \frac{2\pi}{\gamma} f + \Delta H_{low} \left(\frac{H_z}{f}\right)^n \qquad (S1)$$

where the power-law term $f^{-n}$ can describe the linewidth broadening due to the incomplete saturation at low frequency. $\Delta H_{low}$ and $H_z$ are fitting constants. The exponent $n$ are 3 and 5 for magnetic field angle of $\varphi_H = 30°$ and $60°$ gained from the Low-field losses fitting, which are shown in Figures S3(a)-S3(b), respectively. The obtained damping values are summarized in Figure S3(c), which are similar to the linear fitting in the higher frequency range. These results further support the validness of effective Gilbert damping values via linear fitting using only high frequency results, as discussed in the main paper.

The Low-field losses method is also used to fit the effective Gilbert damping of $CrO_2$ along [001] direction. However, the exact values are hard to be determined since multiple values can all well fit the experimental results (Figure S4).

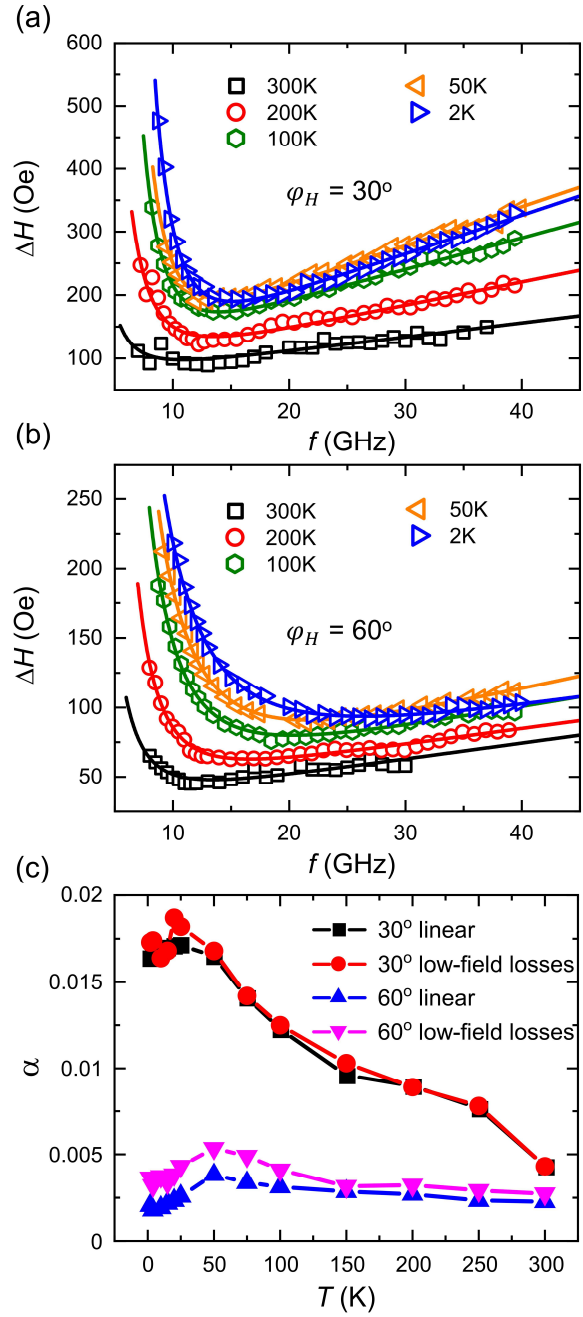

**Figure S3**. Low-field losses fitting. (a, b) Resonance half linewidth as a function of RF frequency from 8 to 40 GHz measured with $\varphi_H = 30°$ and $60°$. The solid lines represent the best fitting curves with eq. (S1). (c) The temperature-dependent effective Gilbert damping obtained from Low-field losses fitting, which are similar to those obtained using linear fitting method.

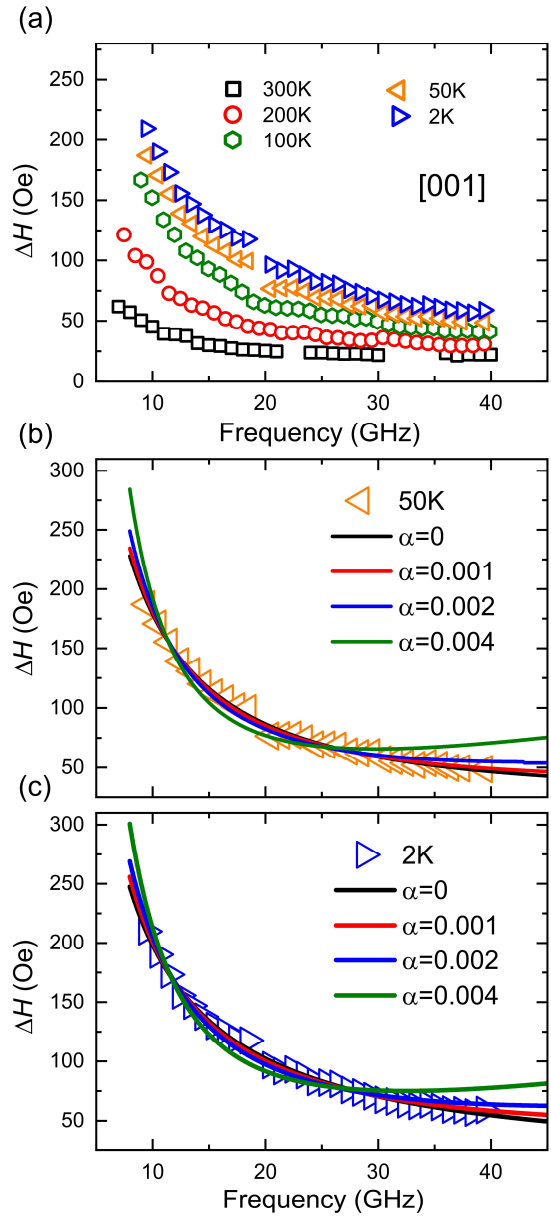

**Figure S4**. Low-filed losses fitting of the effective Gilbert damping along [001]. (a) Half linewidth as a function of RF resonance frequency of [001] direction at various temperatures. (b, c) The failed trials to determine the effective Gilbert damping values at *T* = 50 K and 2 K using eq. (S1).

### III: Supporting data from additional sample

Figure S5 summarizes effective Gilbert damping results obtained from another 170 nm $CrO_2$ thin film. Clearly, a larger effective Gilbert damping is observed when the magnetic field is along the [010] direction, and a smaller damping is observed when the magnetic field is along $\varphi_H = 60°$ direction. As the temperature decreases, the damping along [010] direction increases (Figure S5(a)). While on the other hand, the damping exhibits a decreasing feature below $T = 50$ K. More importantly, the anisotropic feature increases as the temperature decreases (Figure S5(b)). These results, together with the results in the main paper, further supporting extremely large anisotropic effective Gilbert damping in $CrO_2$.

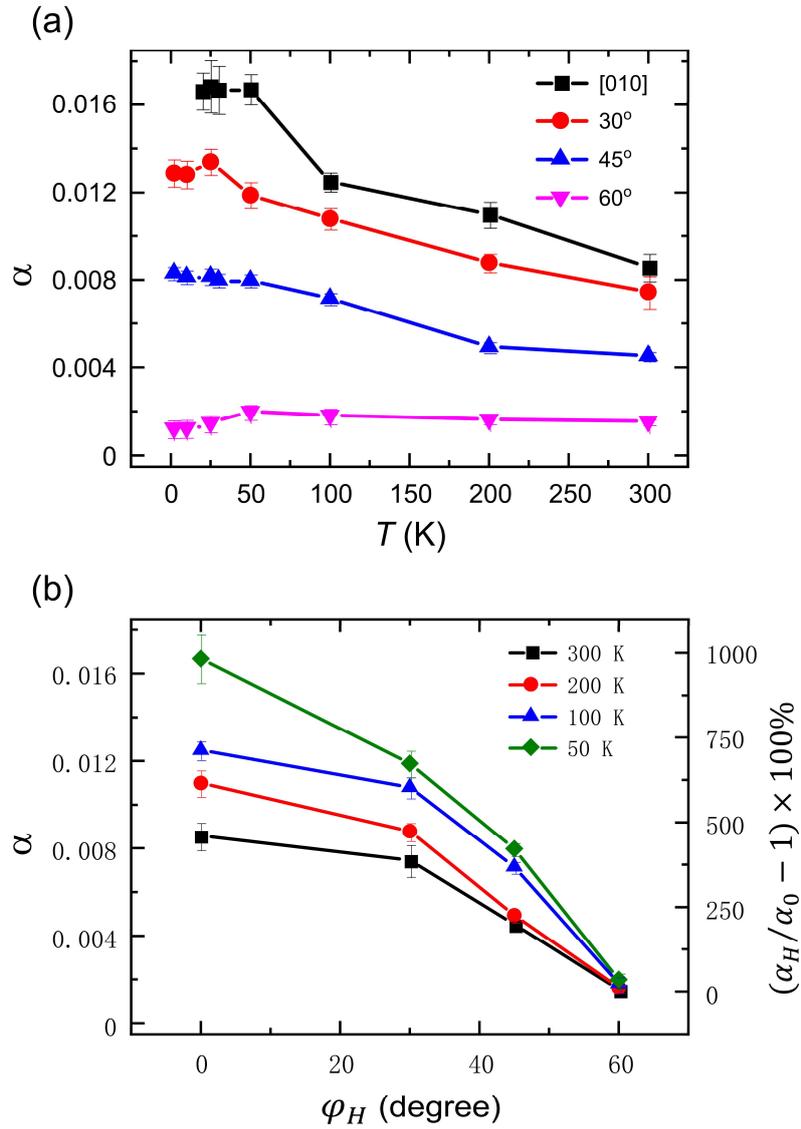

**Figure S5**. Supporting data from another $CrO_2$ thin film. (a) Temperature dependence of the effective Gilbert damping of 170 nm $CrO_2$ film measured for various magnetic field angle, $\varphi_H$. (b) Large anisotropy of effective Gilbert damping in another $CrO_2$ film as a function of $\varphi_H$.

### IV: Exclude the extrinsic effects contributing to the effective Gilbert damping

The extrinsic effect which can broaden the FMR linewidths can be expressed as the following equation[4]

$$\Delta H^{\text{broadening}} = \Delta H^{2M}(f, \varphi_H, \theta_H) + \Delta H^{\text{mosaicity}}(f, \varphi_H, \theta_H) + \Delta H_0(\varphi_H, \theta_H) \quad (S2)$$

Where $\Delta H^{2M}$ and $\Delta H^{\text{mosaicity}}$ are the linewidth comes from two magnon scattering and mosaicity broadening, $\Delta H_0$ is the inhomogeneous term. $f$ is the microwave frequency, $\varphi_H$ is the angle between in-plane magnetic field and hard axis [010] is our measurement, $\theta_H$ is the angle between magnetic field and the normal direction of the $CrO_2$ film which is 90 degrees in our measurement. From our best knowledge, we can first exclude $\Delta H^{2M}$ and $\Delta H_0$'s effect on our conclusion, the diagram of these two terms and our fitted data of $\varphi_H = 60°$ is plotted in Figure S6. The detailed analysis is as following

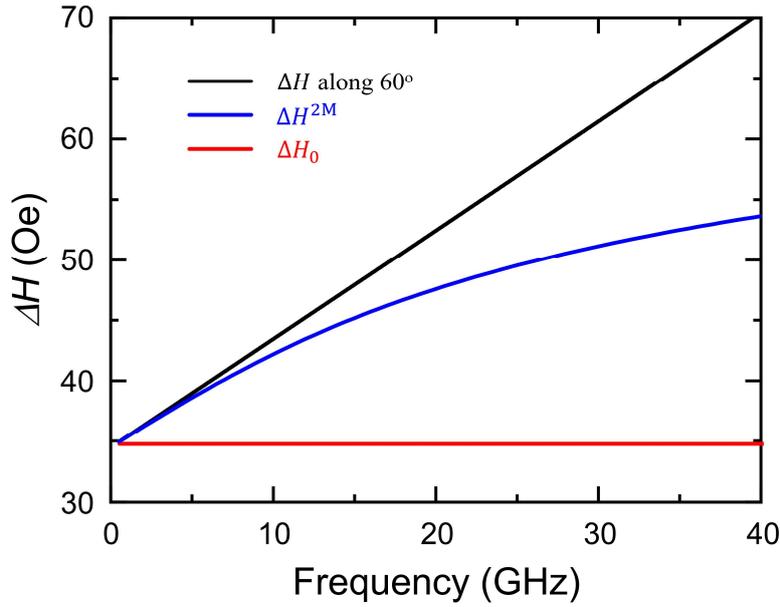

**Figure S6**. Schematic of frequency dependent half linewidth for fitted results at $\varphi_H = 60°$ (black), two magnon scattering term $\Delta H^{2M}$ (blue) and inhomogeneous term $\Delta H_0$ (red).

In our results, the two magnon scattering contribution to the half linewidth can be excluded directly. Because in previous study of two magnon scattering, the $\Delta H^{2M}$ vs. f will have a nonlinear relation[5–7]. However, in our data, Figure 2(a) and Figure 3(a, b), the very good linear relation can promise the two magnon scattering effect could be neglected.

According to some theoretical and experimental works[7–9], $\Delta H_0$ is only dependent on $\varphi_H$ ($\theta_H = 90°$ is fixed). However, in our work like Figure 2(a) and Figure 3(a, b), with the frequency changes, the measured linewidths also change and show great linear relation as a function of frequency, which can exclude $\Delta H_0$'s effect on our observations.

The mosaicity broadening will disappear when the magnetic field is along easy or hard axis[4], which means in our measurement, $\Delta H^{\text{mosaicity}} = 0$ for [010] direction. In Figure 3(a), the halfwidth increases when temperature decreases, which means mosaicity broadening has finite impact on the temperature dependent half linewidths. In Figure S7, we show the halfwidth $\Delta H$ as a function of different $\varphi_H$. The similar behavior of different temperatures could indicate the mosaicity broadening effect is small.

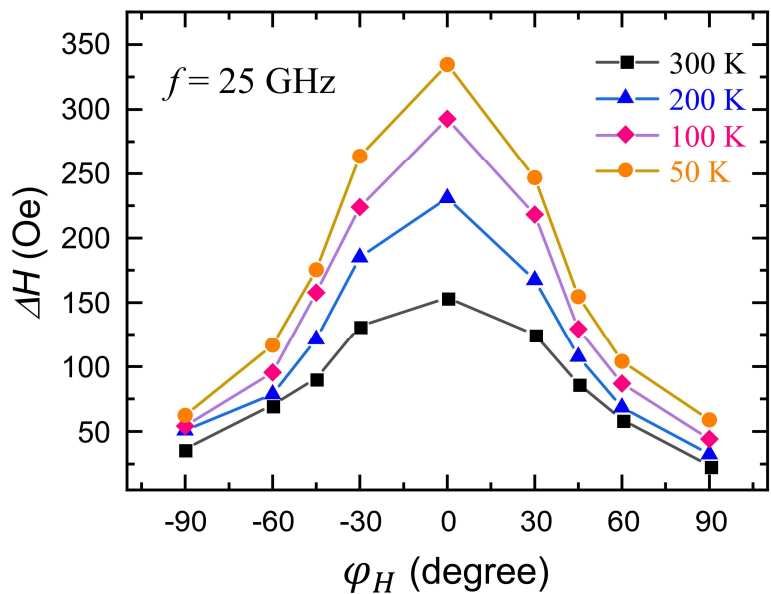

**Figure S7**. The half linewidth $\Delta H$ as a function of external field's direction $\varphi_H$ for different temperatures.

In our work, there are extrinsic effects that contribute to the half linewidth $\Delta H$ and may change with temperature $T$ and magnetic field direction $\varphi_H$. But according to the upper discussion, we believe the extrinsic broadening can be properly excluded for the contribution to our anisotropic and especially temperature-dependent effective Gilbert damping.